\documentclass[twocolumn, showpacs, preprintnumbers, pra, floatfix]{revtex4-1}

\usepackage{graphicx}
\usepackage{epsfig}
\usepackage{bm}
\usepackage{amssymb, amsmath}
\usepackage{verbatim}

\def\4he{$^4$He}
\def\3he{$^3$He}
\def\cm3{cm$^{-3}$}

\begin{document}

\title{Properties of the ground $^3$F$_2$ state and the excited $^3$P$_0$ state of atomic thorium in cold collisions with \3he}

\author{Yat Shan Au,$^{1, 3}$ Colin B. Connolly,$^{1, 3}$ Wolfgang~Ketterle,$^{2, 3}$ and John M. Doyle$^{1, 3}$}
\affiliation{
$^1$Department of Physics, Harvard University, Cambridge, Massachusetts 02138\\
$^2$Department of Physics, Massachusetts Institute of Technology, Cambridge, Massachusetts 02139\\
$^3$Harvard-MIT Center for Ultracold Atoms, Cambridge, Massachusetts 02138
}
\date{\today}

\begin{abstract}
We measure inelastic collisional cross sections for the ground $^3$F$_2$ state and the excited $^3$P$_0$ state of atomic thorium in cold collisions with \3he.  We determine for Th ($^3$F$_2$) at 800~mK the ratio $\gamma \approx 500$ of the momentum-transfer to Zeeman relaxation cross sections for collisions with \3he.  For Th ($^3$P$_0$), we study electronic inelastic processes and find no quenching even after $10^6$ collisions.  We also determine the radiative lifetime of Th ($^3$P$_0$) to be $\tau > 130$~ms.  This great stability of the metastable state opens up the possibility for further study, including trapping.
\end{abstract}



\maketitle

\section*{Introduction}
Ultracold atoms beyond the alkali metals are providing new opportunities in areas as diverse as clocks, physics beyond the standard model, and quantum simulation.  The use of rare earth atoms has enabled the realization of dipolar interactions \cite{Lu2012, Aikawa2012} and the development of new schemes to encode quantum information \cite{Saffman2008}.  Alkaline earth atoms are not only competitive contesters to be the next generation metrological standards \cite{Ludlow2008}, but also enrich the field of cold atomic collisions to include metastable states \cite{Hansen2006, Kokoouline2003}.  As a fundamental platform for understanding these new systems, atom-helium collisions, due to the relatively simple electronic structure of helium, often provide useful links between theory and experiments.  The study of such collisions has led to a better understanding of the roles played by anisotropic electrostatic interactions \cite{Hansen2006, Kokoouline2003, Yamaguchi2008}, spin-orbit coupling \cite{Maxwell2008, Connolly2013} and shielding by outer s-electrons \cite{Lu2009, Connolly2010}.

Anisotropic electrostatic interactions are important for understanding collisions between non-S state atoms and other species, including helium.  During a collision, anisotropic interactions can effectively couple the different projections within an orbital angular momentum $l_m$-manifold, and hence, change the orientation of the magnetic moment.  The ratio $\gamma$ between the momentum transfer and Zeeman relaxation cross section provides a useful metric to characterize such collisions.  Considering benchmark examples of $\gamma$, in the case of anisotropic $^3$P$_2$-state oxygen, $\gamma \sim 7$ \cite{Krems2003}, compared with an isotropic S-state potassium atom $\gamma > 10^8$ \cite{Tscherbul2008}.  The presence of outer s-electrons in a non-S state atom can suppress (``shield'') such electrostatic anisotropy, resulting in $\gamma > 10^4$ \cite{Hancox2004, Hancox2005}.

In this paper, we use atomic thorium (Th) to study the effects of possible shielding in the actinides, the first study of its kind.  In doing so, we extend previous studies of electrostatic anisotropy suppression in transition metals \cite{Hancox2004} and in lanthanides \cite{Hancox2005}, thus providing quantitative comparison of the degree of outer s-electron shielding between 3d4s (e.g. Ti), 4f6s (e.g. Tm) and 6d7s (e.g. Th) systems.

In addition to our study of anistropy in the ground state, we also explore the collisional properties of the first excited $^3$P$_0$ state in thorium.  In contrast to previous studies of metastable $^3$P$_J$ states in alkaline earths \cite{Traverso2009, Kelly1988, Redondo2004}, transitions between fine structure multiplets are not energetically allowed.  This permits a direct study of metastable electronic quenching, and in conjunction, the determination that this metastable state has a very long lifetime, apparently not strongly affected by perturbations in this complex atom.

\section*{Experimental Setup}

The core approach for all of our work is the creation of cold, dilute gases of Th in the presence of cold \3he gas, the collisional partner species.  We prepare cold samples of atomic thorium ($>$ 10$^{11}$) using buffer-gas cooling \cite{Doyle1999}.  Our setup consists of a copper cell at 800~mK, cooled by a dilution refrigerator via a flexible heat link \cite{Johnson2010}.  A pair of superconducting Helmholtz coils creates a uniform magnetic field over the cell region for Zeeman relaxation measurements.  No magnetic field is applied for the excited $^3$P$_0$ state experiments.

Atomic thorium is introduced into the buffer gas via laser ablation of a solid thorium metal target.  Thorium atoms thermalize to the cell temperature via collisions with \3he before diffusing to the cell wall where they stick.  We directly measure the atoms' temperature by fitting optical absorption spectra to a Voigt profile of the 6d$^2$7s$^2$ ($^3$F$_2$) $\rightarrow$ 6d$^2$7s7p ($^3$G$_3$) transition at 372~nm.  We make measurements at late time, at which point the single exponential decay profile of the Th ($^3$F$_2$) optical density indicates that all but the lowest diffusion mode can be ignored \cite{Doyle1999}.  In this case, the diffusion time constant $\tau_d$ in a cylindrical cell of radius $r$ and length $L$ is given by

\begin{equation}
\tau_d = \frac{32}{3\pi}\frac{n_b \sigma_d}{\bar{v}}\left(\frac{j^2_{01}}{r^2} + \frac{\pi^2}{L^2}\right)^{-1},
\end{equation}

\noindent where $n_b$ is the buffer gas density, $\sigma_d$ is the thermally averaged momentum transfer cross section, $j_{01} \approx 2.4$ is the first zero of the Bessel function $J_0$, and $\bar{v}$ is the mean Th-\3he center of mass speed \cite{Hasted1972}.

\section*{Measuring Zeeman Relaxation}

\begin{figure}
\includegraphics[width=8.6 cm]{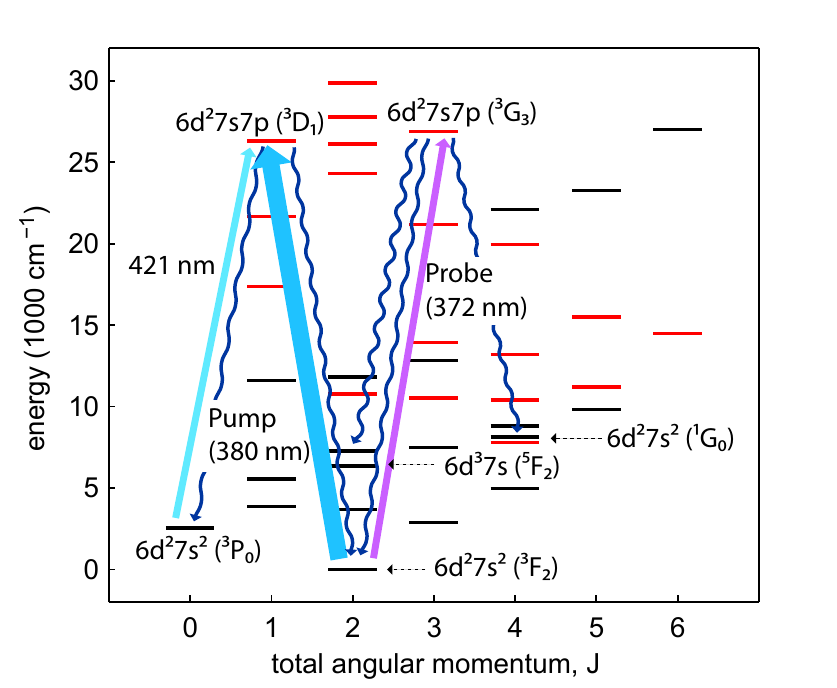}
\caption{\label{fig:level} (Color online) Energy levels of thorium up to 30~000 cm$^{-1}$ \cite{Corliss1962}. Black (red) line indicates parity even (odd) state.  In Zeeman relaxation measurement, the ground $^3$F$_2$ state is pumped by the 6d$^2$7s$^2$ ($^3$F$_2$) $\rightarrow$ 6d$^2$7s7p ($^3$D$_1$) transition at 380~nm and probed by the 6d$^2$7s$^2$ ($^3$F$_2$) $\rightarrow$ 6d$^2$7s7p ($^3$G$_3$) transition at 372~nm.  In the excited $^3$P$_0$ measurements, the state is populated by optical pumping via 6d$^2$7s7p ($^3$D$_1$) and is probed by 6d$^2$7s$^2$ ($^3$P$_0$) $\rightarrow$ 6d$^2$7s7p ($^3$D$_1$) transition at 421~nm.}
\end{figure}

To measure Zeeman relaxation, we apply a uniform magnetic field of  up to 2~T to spectroscopically resolve the Zeeman sublevels (m$_J$) of the thorium ground $^3$F$_2$ state, which we probe using absorption spectroscopy on the 372-nm transition.  We drive the system away from thermal equilibrium and then monitor the repopulation from other Zeeman sublevels via inelastic collisions.  Here we deplete a high-field-seeking (HFS) $m_J > 0$ state via an optical pumping pulse using the 6d$^2$7s$^2$ ($^3$F$_2$) $\rightarrow$ 6d$^2$7s7p ($^3$D$_1$) transition at 380~nm, as shown in Fig. \ref{fig:level}.

The expected Zeeman shift of an atomic transition can be calculated from the Land\'{e} g-factors, $g_J$, of the terms of the lower and the upper states using Eq. \ref{eqn:g_term}.

\begin{equation}
g_J = \frac{3}{2} + \frac{S(S + 1) - L(L + 1)}{2J(J+1)}
\label{eqn:g_term}
\end{equation}

\noindent if both $L$ and $S$ are good quantum numbers for the states.  The literature value for Th ($^3$F$_2$) is $g_J=0.736$ \cite{Katz1957}, compared to $g_J=2/3$ predicted by Eq.~\ref{eqn:g_term}.  The discrepancy is not unexpected for heavy atoms, and it suggests a significant relativistic perturbation to the atomic states, which can be of importance to the cold collisional properties \cite{Maxwell2008}.  We adopt the literature value for the ground state $g_J$, and account for the difference as observed values for the excited states $g_J$ (Table \ref{tab:g_factor}).

\begin{table}
\begin{ruledtabular}
\begin{tabular}{ccccc}
level (cm$^{-1}$) &term & $g_J$ (calculated) & $g_J$ (observed)\\
\hline
0 & $^3$F$_2$ & 0.667 & 0.736 \cite{Katz1957}\\
26 287.049	 & $^3$D$_1$ & 0.5 & $0.9 \pm 0.05$ (this work)\\
26 878.162 & $^3$G$_3$ & 0.75 & $1.1 \pm 0.05$ (this work)\\
\end{tabular}
\end{ruledtabular}
\caption{\label{tab:g_factor} Some of the measured g-factors that are different from their values calculated using spectroscopic term assignments (Eq.~\ref{eqn:g_term}).  We use the literature value \cite{Katz1957} for the ground state's g-factor and deduce g-factors of the excited states from the observed Zeeman shifts.  The frequency shifts were measured using a wavemeter.  The uncertainty in $g_J$ is dominated by current-to-field calibration of the coils.}
\end{table}

The method used here for measuring the cross section ratio $\gamma$, i.e. by observing the recovery of HFS states, is different from that used in \cite{Johnson2010}, which relies on monitoring the decay of a low-field-seeking (LFS) state populated during the initial laser ablation.  In the latter case,  measurements made soon after ablation are complicated by high-order diffusion modes and thermal effects from the laser ablation.  Relying only on late-time data to avoid these effects limits the sensitivity of that method to $\gamma \gtrsim 1000$ \cite{Connolly2012}.
  
\begin{figure}
\includegraphics[width=8.6 cm]{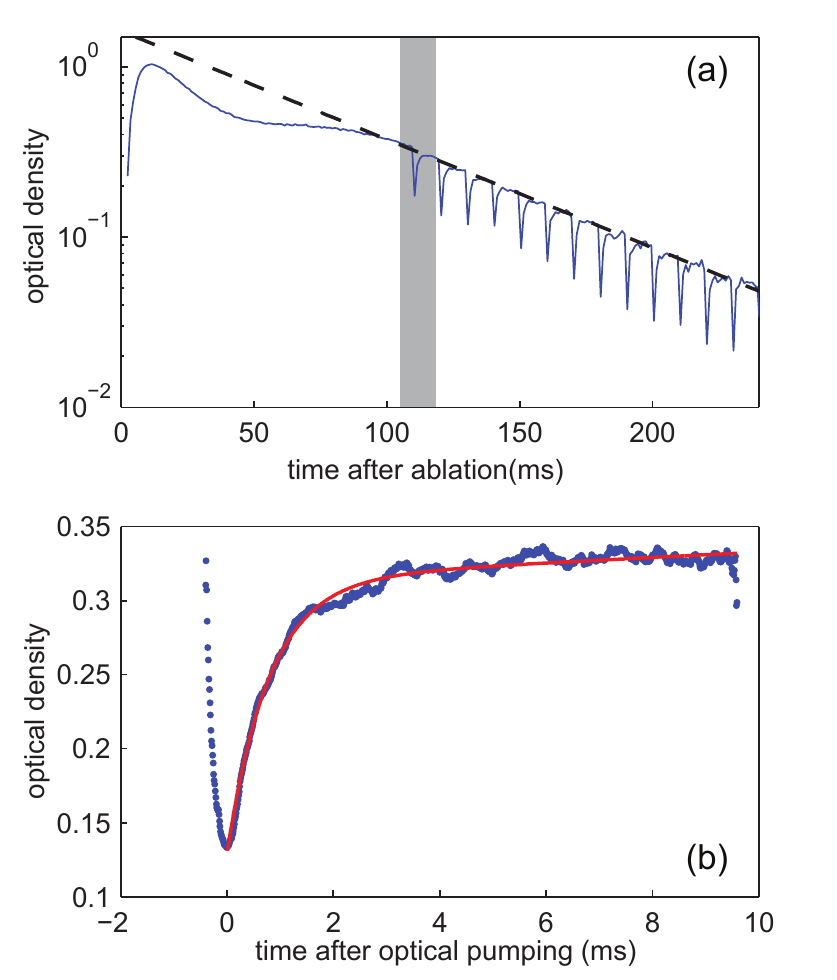}
\caption{\label{fig:Th_Zeeman_data}  (Color online) (a) The optical density (OD) of atomic thorium during a Zeeman relaxation measurement.  Measurements are performed starting 100~ms after ablation, when both diffusion modes and temperature are stabilized, as verified by a simple exponential decay of the atomic OD (dashed-line).  (b), corresponding to the shaded area in (a) after subtraction of the diffusive baseline, shows the depletion of OD by optical pumping and recovery of OD via inelastic collisions.  Multiple measurements are made with separate optical pumping pulses before the atoms diffuse to the cell wall.}
\end{figure}

In this work, the use of optical pumping to drive the system out of equilibrium allows us to perform the measurements after the decay of high-order diffusion modes and the establishment of thermal stabilization, while preserving sensitivity to short time scales.  The observation of a simple exponential decay of the atomic optical density during the measurements confirms that no other processes interfere with simple Zeeman relaxation (Fig.~\ref{fig:Th_Zeeman_data}).  This method thus theoretically enables a measurement of $\gamma$ to values as low as $\approx 1$.  Additionally, using optical pumping not only allows a higher data rate by enabling repetition of relaxation measurements within an atom production (see Fig. \ref{fig:Th_Zeeman_data}), but also provides a robust systematic check if we ensure the measured value is independent of time delay from initial atom production \cite{Lu2009, Connolly2013}.

The observed time constant $\tau_{Z}$ for the repopulation of the HFS state via inelastic collisions is given by \cite{Connolly2013b}

\begin{equation}
\label{Eq:gamma_fit}
\tau_Z = \frac{\gamma}{{\bar{v}}^2G\tau_d}
\end{equation}

\begin{figure}
\includegraphics[width=8.6 cm]{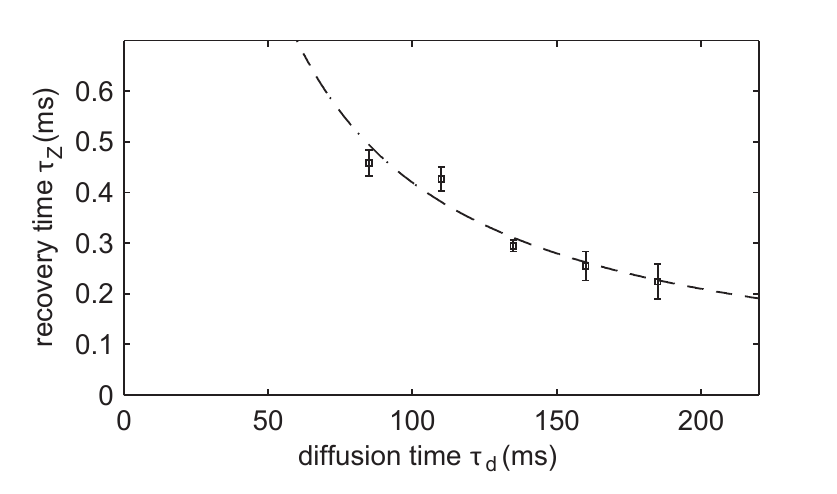}
\caption{\label{fig:tau_Z_tau_d}  Zeeman relaxation measurement.  Fitted repopulation time $\tau_{Z}$ of the most high-field seeking (HFS) state ($m_J = J$) of atomic thorium at different $^3$He densities ($\tau_d \propto n_\mathrm{He}$) at 0.5~T.  Error bars are statistical uncertainties.  The ratio $\gamma$ of the momentum-transfer to Zeeman relaxation cross section is given by fitting to Eq.~\ref{Eq:gamma_fit}.}
\end{figure}

\begin{figure}
\includegraphics[width=8.6 cm]{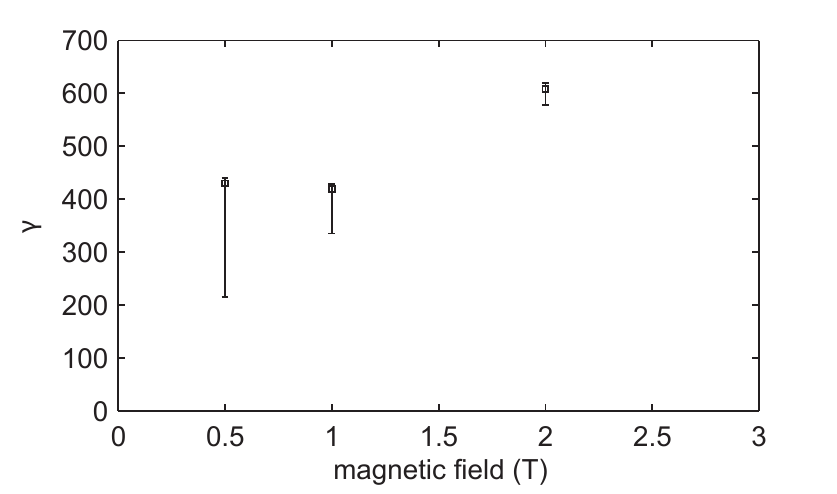}
\caption{\label{fig:gamma_B} Momentum-transfer-to-Zeeman-relaxation collision ratio $\gamma$ for Th ($^3$F$_2$) ground state measured at 800~mK.}
\end{figure}

As shown in Fig. \ref{fig:tau_Z_tau_d},  the measured dependence of the recovery time of the HFS state on its diffusion time agrees with Eq.\ref{Eq:gamma_fit}, where the value for $\gamma$ can be extracted as a fit parameter.

At finite temperature, an inelastic collision can sometimes have sufficient energy to promote a thorium atom from the stretched HFS state to higher sublevels, slowing the relaxation to equilibrium and resulting in an overestimation of $\gamma$.  We simulated our system using the model developed in \cite{Johnson2010}.  In Fig. \ref{fig:gamma_B}, the upper limit of the error bars is the statistical uncertainty from fitting, while the lower limit represents the uncertainty due to such thermal excitations.

We also repeat the Zeeman relaxation measurement on the ground-state sublevel with $\mathrm{m_J = J -1}$ at 0.5~T, and found $\gamma$ to be statistically similar to that of the most HFS state ($\mathrm{m_J = J}$).  We could not make this comparison at higher magnetic field due to insufficient thermal population of the $\mathrm{m_J = J -1}$ state.

A source of possible measurement error is the filling of the HFS state from long-lived states (e.g. $^5$F$_2$.  see Fig.\ref{fig:level}) populated during optical pumping due to nonzero branching from the excited $^3$D$_1$ state.  The state lifetimes and transition strengths are not sufficiently known \cite{Corliss1962} to quantify the populations of other excited states or the rate of decay into the ground state.  However, the populations in such states due to optical pumping will depend sensitively on both power and duration of the pump and probe lasers.  We change these parameters by a factor of 2 and obtain a statistically equivalent values of $\gamma$.  Therefore we conclude that decay from other states does not affect our measurement.

\section*{Excited $^3$P$_0$ State}

\begin{figure}
\includegraphics[width=8.6 cm]{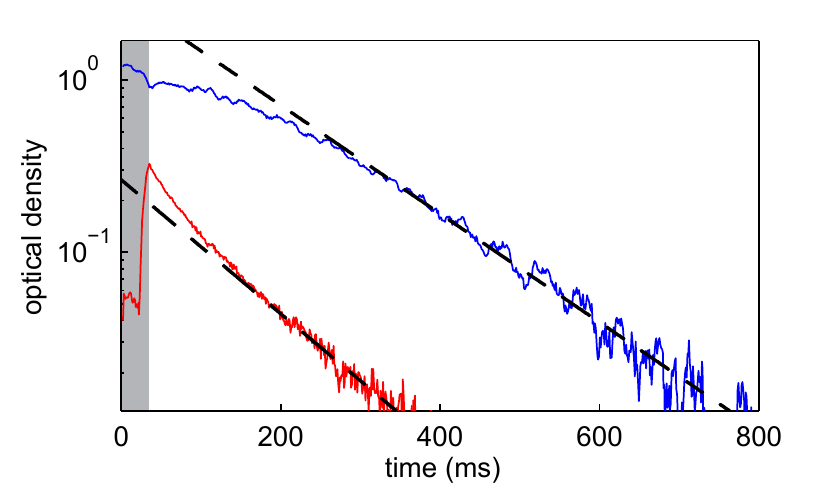}
\caption{\label{fig:Th_P_data}  (Color online)  The optical density (OD) of atomic thorium during the excited  $^3$P$_0$ state measurement.  Blue (red) is the OD of the $^3$F$_2$ ($^3$P$_0$) state.  Optical pumping is applied in the first 20~ms (shaded area) to transfer population from the ground  $^3$F$_2$ state to the excited $^3$P$_0$ state.  Dashed lines are single-exponential fits to the lowest diffusion mode at late times.  Diffusive fits start at 310~ms and 180~ms for the $^3$P$_0$ and the $^3$F$_2$ state respectively.}
\end{figure}

We also apply the optical pump and probe technique to study collisions in the metastable first excited state of atomic thorium, the $^3$P$_0$ state.  The state is probed using the 6d$^2$7s$^2$ ($^3$P$_0$) $\rightarrow$ 6d$^2$7s7p ($^3$D$_1$) transition at 421~nm.  When the cold thorium gas is produced via laser ablation, a small fraction (optical density $\sim$ 0.01) is populated in the $^3$P$_0$ state.  We increase the state population by an order of magnitude using optical pumping via 6d$^2$7s7p ($^3$D$_1$) (see Fig.~\ref{fig:level} and Fig.~\ref{fig:Th_P_data}).  We measure the relaxation time of the $^3$P$_0$ state at various buffer gas densities to search for collisional quenching of the excited state.

\begin{figure} 
\includegraphics[width=8.6 cm]{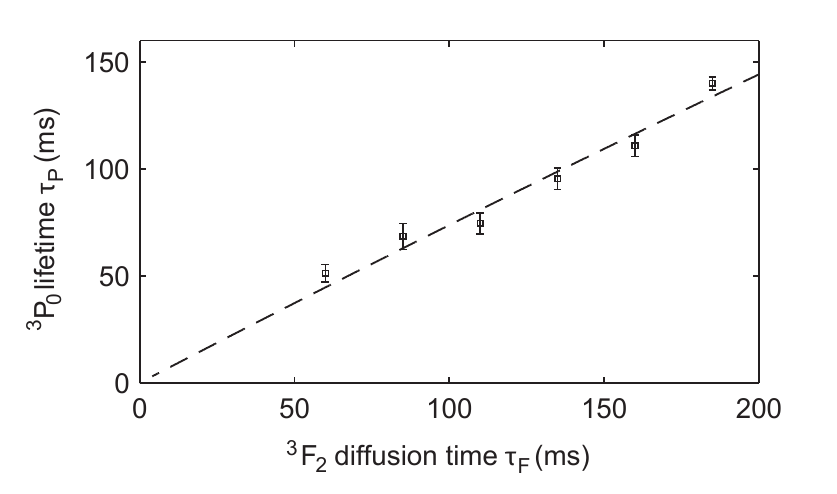}
\caption{\label{fig:Th_P_fit}  Linear dependence of the excited $^3$P$_0$ state lifetime $\tau_P$ on diffusion time and buffer gas density.}
\end{figure}

The results for the $^3$P$_0$ state are shown in Fig.~\ref{fig:Th_P_fit}.  The excited state lifetime $\tau_P$ shows a linear dependence on the ground state ($^3$F$_2$) diffusion time $\tau_F$, which is proportional to the buffer gas density.  The lifetime $\tau_P$ is determined by five effects: refilling from a reservoir, optical pumping, diffusion, collisional quenching and radiative decay.

We first consider the possibility that the $^3$P$_0$ state is constantly refilled by populations from reservoir states with higher energies, the only mechanism of the four that can increase the observed value of $\tau_P$.  We probe the $^3$P$_0$ state with sufficient laser power so that $\tau_P$ is determined by optical pumping by the probe light instead of diffusion or radiative decay.  We shutter the probe beam for 100~ms before probing the state again.  We do not observe a recover of the $^3$P$_0$ population, and thus we conclude that contribution to $\tau_P$ by refilling from a reservoir is insignificant.

Detector efficiency limits our probe laser power $>~150$~nW.  We observe a negatively correlated power dependent slope.  We do not have sufficient data to eliminate the effect on $\tau_P$ imposed by optical pumping by extrapolating to zero probe power.

The laser-independent component of the slope (in Fig.~\ref{fig:Th_P_fit}) is determined by the ratio for the elastic cross-section between Th ($^3$P$_0$)-\3he and Th ($^3$F$_2$)-\3he (i.e. diffusion).

Finally, collisional quenching causes a negative slope and radiative decay imposes a $^3$He density ($\tau_F \propto n_\mathrm{He}$) cap on $\tau_P$.  Although we do not separate the contribution to the slope by optical pumping and diffusion, the linear dependence (positively sloped) of $\tau_P$ on $\tau_F$ enable us to conclude that no electronic quenching collisions were observed.  Using the highest buffer gas density data point, we bound the value of $\gamma > 10^6$ and the radiative lifetime of the state $\tau_R > 130$~ms.

\section*{Conclusion}

We measure the ratio $\gamma$ of the momentum-transfer to the magnetic reorientation cross sections of Th in the ground $^3$F$_2$ state.  We find that $\gamma \approx 500$, which is well above that of open-shell oxygen, but well below that of submerged-shell lanthanides.  This value of $\gamma$ is too small for direct buffer gas loading into a magnetic trap \cite{Decarvalho2004}, but it is sufficient for indirect loading from a buffer gas beam \cite{Lu2011}.  This brings the possibility of studying trapped, cold atomic thorium, and perhaps opening up the actinides to cold atom physics. 

We also study metastable Th, setting a lower bound of $\gamma > 10^6$ for electronic quenching of the first excited $^3$P$_0$ state.  This very low quenching cross section allows for the determination of a long radiative lifetime for Th ($^3$P$_0$) of $\tau >130$~ms. This invites the possibility of studying ultracold metastable Th in an optical lattice, as was done with the lanthanide Yb \cite{Yamaguchi2008}.

Together with the unexplained results of collisions between Yb atoms in $^3$P$_0$ states \cite{Yamaguchi2008}, there may be a general trend of robustness in the excited $^3$P$_0$ state, opening up new opportunities for applications using metastable atoms.

\begin{acknowledgments}

We acknowledge Timur Tscherbul for helpful theoretical inputs and Elizabeth Petrik for assistance in target preparation.  This work was supported by the NSF through the Harvard-MIT Center for Ultracold Atoms.

\end{acknowledgments}

\bibliographystyle{apsrev}
\bibliography{Th_PRA_05}

\begin{thebibliography}{27}
\expandafter\ifx\csname natexlab\endcsname\relax\def\natexlab#1{#1}\fi
\expandafter\ifx\csname bibnamefont\endcsname\relax
  \def\bibnamefont#1{#1}\fi
\expandafter\ifx\csname bibfnamefont\endcsname\relax
  \def\bibfnamefont#1{#1}\fi
\expandafter\ifx\csname citenamefont\endcsname\relax
  \def\citenamefont#1{#1}\fi
\expandafter\ifx\csname url\endcsname\relax
  \def\url#1{\texttt{#1}}\fi
\expandafter\ifx\csname urlprefix\endcsname\relax\def\urlprefix{URL }\fi
\providecommand{\bibinfo}[2]{#2}
\providecommand{\eprint}[2][]{\url{#2}}

\bibitem[{\citenamefont{Lu et~al.}(2012)\citenamefont{Lu, Burdick, and
  Lev}}]{Lu2012}
\bibinfo{author}{\bibfnamefont{M.}~\bibnamefont{Lu}},
  \bibinfo{author}{\bibfnamefont{N.~Q.} \bibnamefont{Burdick}},
  \bibnamefont{and} \bibinfo{author}{\bibfnamefont{B.~L.} \bibnamefont{Lev}},
  \bibinfo{journal}{Phys. Rev. Lett.} \textbf{\bibinfo{volume}{108}},
  \bibinfo{pages}{215301} (\bibinfo{year}{2012}).

\bibitem[{\citenamefont{Aikawa et~al.}(2012)\citenamefont{Aikawa, Frisch, Mark,
  Baier, Rietzler, Grimm, and Ferlaino}}]{Aikawa2012}
\bibinfo{author}{\bibfnamefont{K.}~\bibnamefont{Aikawa}},
  \bibinfo{author}{\bibfnamefont{A.}~\bibnamefont{Frisch}},
  \bibinfo{author}{\bibfnamefont{M.}~\bibnamefont{Mark}},
  \bibinfo{author}{\bibfnamefont{S.}~\bibnamefont{Baier}},
  \bibinfo{author}{\bibfnamefont{A.}~\bibnamefont{Rietzler}},
  \bibinfo{author}{\bibfnamefont{R.}~\bibnamefont{Grimm}}, \bibnamefont{and}
  \bibinfo{author}{\bibfnamefont{F.}~\bibnamefont{Ferlaino}},
  \bibinfo{journal}{Phys. Rev. Lett.} \textbf{\bibinfo{volume}{108}},
  \bibinfo{pages}{210401} (\bibinfo{year}{2012}).

\bibitem[{\citenamefont{Saffman and Molmer}(2008)}]{Saffman2008}
\bibinfo{author}{\bibfnamefont{M.}~\bibnamefont{Saffman}} \bibnamefont{and}
  \bibinfo{author}{\bibfnamefont{K.}~\bibnamefont{Molmer}},
  \bibinfo{journal}{Phys. Rev. A} \textbf{\bibinfo{volume}{78}},
  \bibinfo{pages}{012336} (\bibinfo{year}{2008}).

\bibitem[{\citenamefont{Ludlow et~al.}(2008)\citenamefont{Ludlow, Zelevinsky,
  Campbell, Blatt, Boyd, de~Miranda, Martin, Thomsen, Foreman, Ye
  et~al.}}]{Ludlow2008}
\bibinfo{author}{\bibfnamefont{A.~D.} \bibnamefont{Ludlow}},
  \bibinfo{author}{\bibfnamefont{T.}~\bibnamefont{Zelevinsky}},
  \bibinfo{author}{\bibfnamefont{G.~K.} \bibnamefont{Campbell}},
  \bibinfo{author}{\bibfnamefont{S.}~\bibnamefont{Blatt}},
  \bibinfo{author}{\bibfnamefont{M.~M.} \bibnamefont{Boyd}},
  \bibinfo{author}{\bibfnamefont{M.~H.~G.} \bibnamefont{de~Miranda}},
  \bibinfo{author}{\bibfnamefont{M.~J.} \bibnamefont{Martin}},
  \bibinfo{author}{\bibfnamefont{J.~W.} \bibnamefont{Thomsen}},
  \bibinfo{author}{\bibfnamefont{S.~M.} \bibnamefont{Foreman}},
  \bibinfo{author}{\bibfnamefont{J.}~\bibnamefont{Ye}}, \bibnamefont{et~al.},
  \bibinfo{journal}{Science} \textbf{\bibinfo{volume}{319}},
  \bibinfo{pages}{1805} (\bibinfo{year}{2008}).

\bibitem[{\citenamefont{Hansen and Hemmerich}(2006)}]{Hansen2006}
\bibinfo{author}{\bibfnamefont{D.}~\bibnamefont{Hansen}} \bibnamefont{and}
  \bibinfo{author}{\bibfnamefont{A.}~\bibnamefont{Hemmerich}},
  \bibinfo{journal}{Phys. Rev. Lett.} \textbf{\bibinfo{volume}{96}},
  \bibinfo{pages}{073003} (\bibinfo{year}{2006}).

\bibitem[{\citenamefont{Kokoouline et~al.}(2003)\citenamefont{Kokoouline,
  Santra, and Greene}}]{Kokoouline2003}
\bibinfo{author}{\bibfnamefont{V.}~\bibnamefont{Kokoouline}},
  \bibinfo{author}{\bibfnamefont{R.}~\bibnamefont{Santra}}, \bibnamefont{and}
  \bibinfo{author}{\bibfnamefont{C.}~\bibnamefont{Greene}},
  \bibinfo{journal}{Phys. Rev. Lett.} \textbf{\bibinfo{volume}{90}},
  \bibinfo{pages}{253201} (\bibinfo{year}{2003}).

\bibitem[{\citenamefont{Yamaguchi et~al.}(2008)\citenamefont{Yamaguchi, Uetake,
  Hashimoto, Doyle, and Takahashi}}]{Yamaguchi2008}
\bibinfo{author}{\bibfnamefont{A.}~\bibnamefont{Yamaguchi}},
  \bibinfo{author}{\bibfnamefont{S.}~\bibnamefont{Uetake}},
  \bibinfo{author}{\bibfnamefont{D.}~\bibnamefont{Hashimoto}},
  \bibinfo{author}{\bibfnamefont{J.}~\bibnamefont{Doyle}}, \bibnamefont{and}
  \bibinfo{author}{\bibfnamefont{Y.}~\bibnamefont{Takahashi}},
  \bibinfo{journal}{Phys. Rev. Lett.} \textbf{\bibinfo{volume}{101}},
  \bibinfo{pages}{233002} (\bibinfo{year}{2008}).

\bibitem[{\citenamefont{Maxwell et~al.}(2008)\citenamefont{Maxwell, Hummon,
  Wang, Buchachenko, Krems, and Doyle}}]{Maxwell2008}
\bibinfo{author}{\bibfnamefont{S.}~\bibnamefont{Maxwell}},
  \bibinfo{author}{\bibfnamefont{M.}~\bibnamefont{Hummon}},
  \bibinfo{author}{\bibfnamefont{Y.}~\bibnamefont{Wang}},
  \bibinfo{author}{\bibfnamefont{A.}~\bibnamefont{Buchachenko}},
  \bibinfo{author}{\bibfnamefont{R.}~\bibnamefont{Krems}}, \bibnamefont{and}
  \bibinfo{author}{\bibfnamefont{J.}~\bibnamefont{Doyle}},
  \bibinfo{journal}{Phys. Rev. A} \textbf{\bibinfo{volume}{78}},
  \bibinfo{pages}{042706} (\bibinfo{year}{2008}).

\bibitem[{\citenamefont{Connolly
  et~al.}(2013{\natexlab{a}})\citenamefont{Connolly, Au, Chae, Tscherbul,
  Buchachenko, Lu, Ketterle, and Doyle}}]{Connolly2013}
\bibinfo{author}{\bibfnamefont{C.~B.} \bibnamefont{Connolly}},
  \bibinfo{author}{\bibfnamefont{Y.~S.} \bibnamefont{Au}},
  \bibinfo{author}{\bibfnamefont{E.}~\bibnamefont{Chae}},
  \bibinfo{author}{\bibfnamefont{T.~V.} \bibnamefont{Tscherbul}},
  \bibinfo{author}{\bibfnamefont{A.~A.} \bibnamefont{Buchachenko}},
  \bibinfo{author}{\bibfnamefont{H.-I.} \bibnamefont{Lu}},
  \bibinfo{author}{\bibfnamefont{W.}~\bibnamefont{Ketterle}}, \bibnamefont{and}
  \bibinfo{author}{\bibfnamefont{J.~M.} \bibnamefont{Doyle}},
  \bibinfo{journal}{Phys. Rev. Lett.} \textbf{\bibinfo{volume}{110}},
  \bibinfo{pages}{173202} (\bibinfo{year}{2013}{\natexlab{a}}).

\bibitem[{\citenamefont{Lu et~al.}(2009)\citenamefont{Lu, Singh, and
  Weinstein}}]{Lu2009}
\bibinfo{author}{\bibfnamefont{M.-J.} \bibnamefont{Lu}},
  \bibinfo{author}{\bibfnamefont{V.}~\bibnamefont{Singh}}, \bibnamefont{and}
  \bibinfo{author}{\bibfnamefont{J.}~\bibnamefont{Weinstein}},
  \bibinfo{journal}{Phys. Rev. A} \textbf{\bibinfo{volume}{79}},
  \bibinfo{pages}{050702} (\bibinfo{year}{2009}).

\bibitem[{\citenamefont{Connolly et~al.}(2010)\citenamefont{Connolly, Au,
  Doret, Ketterle, and Doyle}}]{Connolly2010}
\bibinfo{author}{\bibfnamefont{C.~B.} \bibnamefont{Connolly}},
  \bibinfo{author}{\bibfnamefont{Y.~S.} \bibnamefont{Au}},
  \bibinfo{author}{\bibfnamefont{S.~C.} \bibnamefont{Doret}},
  \bibinfo{author}{\bibfnamefont{W.}~\bibnamefont{Ketterle}}, \bibnamefont{and}
  \bibinfo{author}{\bibfnamefont{J.~M.} \bibnamefont{Doyle}},
  \bibinfo{journal}{Phys. Rev. A} \textbf{\bibinfo{volume}{81}},
  \bibinfo{pages}{010702} (\bibinfo{year}{2010}).

\bibitem[{\citenamefont{Krems and Dalgarno}(2003)}]{Krems2003}
\bibinfo{author}{\bibfnamefont{R.}~\bibnamefont{Krems}} \bibnamefont{and}
  \bibinfo{author}{\bibfnamefont{A.}~\bibnamefont{Dalgarno}},
  \bibinfo{journal}{Phys. Rev. A} \textbf{\bibinfo{volume}{68}},
  \bibinfo{pages}{013406} (\bibinfo{year}{2003}).

\bibitem[{\citenamefont{Tscherbul et~al.}(2008)\citenamefont{Tscherbul, Zhang,
  Sadeghpour, Dalgarno, Brahms, Au, and Doyle}}]{Tscherbul2008}
\bibinfo{author}{\bibfnamefont{T.}~\bibnamefont{Tscherbul}},
  \bibinfo{author}{\bibfnamefont{P.}~\bibnamefont{Zhang}},
  \bibinfo{author}{\bibfnamefont{H.}~\bibnamefont{Sadeghpour}},
  \bibinfo{author}{\bibfnamefont{a.}~\bibnamefont{Dalgarno}},
  \bibinfo{author}{\bibfnamefont{N.}~\bibnamefont{Brahms}},
  \bibinfo{author}{\bibfnamefont{Y.~S.} \bibnamefont{Au}}, \bibnamefont{and}
  \bibinfo{author}{\bibfnamefont{J.}~\bibnamefont{Doyle}},
  \bibinfo{journal}{Phys. Rev. A} \textbf{\bibinfo{volume}{78}},
  \bibinfo{pages}{060703} (\bibinfo{year}{2008}).

\bibitem[{\citenamefont{Hancox et~al.}(2004)\citenamefont{Hancox, Doret,
  Hummon, Luo, and Doyle}}]{Hancox2004}
\bibinfo{author}{\bibfnamefont{C.~I.} \bibnamefont{Hancox}},
  \bibinfo{author}{\bibfnamefont{S.~C.} \bibnamefont{Doret}},
  \bibinfo{author}{\bibfnamefont{M.~T.} \bibnamefont{Hummon}},
  \bibinfo{author}{\bibfnamefont{L.}~\bibnamefont{Luo}}, \bibnamefont{and}
  \bibinfo{author}{\bibfnamefont{J.~M.} \bibnamefont{Doyle}},
  \bibinfo{journal}{Nature} \textbf{\bibinfo{volume}{431}},
  \bibinfo{pages}{281} (\bibinfo{year}{2004}).

\bibitem[{\citenamefont{Hancox et~al.}(2005)\citenamefont{Hancox, Doret,
  Hummon, Krems, and Doyle}}]{Hancox2005}
\bibinfo{author}{\bibfnamefont{C.}~\bibnamefont{Hancox}},
  \bibinfo{author}{\bibfnamefont{S.}~\bibnamefont{Doret}},
  \bibinfo{author}{\bibfnamefont{M.}~\bibnamefont{Hummon}},
  \bibinfo{author}{\bibfnamefont{R.}~\bibnamefont{Krems}}, \bibnamefont{and}
  \bibinfo{author}{\bibfnamefont{J.}~\bibnamefont{Doyle}},
  \bibinfo{journal}{Phys. Rev. Lett.} \textbf{\bibinfo{volume}{94}},
  \bibinfo{pages}{013201} (\bibinfo{year}{2005}).

\bibitem[{\citenamefont{Traverso et~al.}(2009)\citenamefont{Traverso,
  Chakraborty, {Martinez de Escobar}, Mickelson, Nagel, Yan, and
  Killian}}]{Traverso2009}
\bibinfo{author}{\bibfnamefont{a.}~\bibnamefont{Traverso}},
  \bibinfo{author}{\bibfnamefont{R.}~\bibnamefont{Chakraborty}},
  \bibinfo{author}{\bibfnamefont{Y.}~\bibnamefont{{Martinez de Escobar}}},
  \bibinfo{author}{\bibfnamefont{P.}~\bibnamefont{Mickelson}},
  \bibinfo{author}{\bibfnamefont{S.}~\bibnamefont{Nagel}},
  \bibinfo{author}{\bibfnamefont{M.}~\bibnamefont{Yan}}, \bibnamefont{and}
  \bibinfo{author}{\bibfnamefont{T.}~\bibnamefont{Killian}},
  \bibinfo{journal}{Phys. Rev. A} \textbf{\bibinfo{volume}{79}},
  \bibinfo{pages}{060702} (\bibinfo{year}{2009}).

\bibitem[{\citenamefont{Kelly et~al.}(1988)\citenamefont{Kelly, Harris, and
  Gallagher}}]{Kelly1988}
\bibinfo{author}{\bibfnamefont{J.~F.} \bibnamefont{Kelly}},
  \bibinfo{author}{\bibfnamefont{M.}~\bibnamefont{Harris}}, \bibnamefont{and}
  \bibinfo{author}{\bibfnamefont{A.}~\bibnamefont{Gallagher}},
  \bibinfo{journal}{Phys. Rev. A} \textbf{\bibinfo{volume}{37}},
  \bibinfo{pages}{2354} (\bibinfo{year}{1988}).

\bibitem[{\citenamefont{Redondo et~al.}(2004)\citenamefont{Redondo,
  {S\'{a}nchez Rayo}, Ecija, Husain, and Casta\~{n}o}}]{Redondo2004}
\bibinfo{author}{\bibfnamefont{C.}~\bibnamefont{Redondo}},
  \bibinfo{author}{\bibfnamefont{M.}~\bibnamefont{{S\'{a}nchez Rayo}}},
  \bibinfo{author}{\bibfnamefont{P.}~\bibnamefont{Ecija}},
  \bibinfo{author}{\bibfnamefont{D.}~\bibnamefont{Husain}}, \bibnamefont{and}
  \bibinfo{author}{\bibfnamefont{F.}~\bibnamefont{Casta\~{n}o}},
  \bibinfo{journal}{Chem. Phys. Lett.} \textbf{\bibinfo{volume}{392}},
  \bibinfo{pages}{116} (\bibinfo{year}{2004}).

\bibitem[{\citenamefont{Doyle et~al.}(1999)\citenamefont{Doyle, Friedrich,
  Guillet, Kim, Patterson, and Weinstein}}]{Doyle1999}
\bibinfo{author}{\bibfnamefont{J.~M.} \bibnamefont{Doyle}},
  \bibinfo{author}{\bibfnamefont{B.}~\bibnamefont{Friedrich}},
  \bibinfo{author}{\bibfnamefont{T.}~\bibnamefont{Guillet}},
  \bibinfo{author}{\bibfnamefont{J.}~\bibnamefont{Kim}},
  \bibinfo{author}{\bibfnamefont{D.}~\bibnamefont{Patterson}},
  \bibnamefont{and} \bibinfo{author}{\bibfnamefont{J.~D.}
  \bibnamefont{Weinstein}}, \textbf{\bibinfo{volume}{309}},
  \bibinfo{pages}{289} (\bibinfo{year}{1999}).

\bibitem[{\citenamefont{Johnson et~al.}(2010)\citenamefont{Johnson, Newman,
  Brahms, Doyle, Kleppner, and Greytak}}]{Johnson2010}
\bibinfo{author}{\bibfnamefont{C.}~\bibnamefont{Johnson}},
  \bibinfo{author}{\bibfnamefont{B.}~\bibnamefont{Newman}},
  \bibinfo{author}{\bibfnamefont{N.}~\bibnamefont{Brahms}},
  \bibinfo{author}{\bibfnamefont{J.~M.} \bibnamefont{Doyle}},
  \bibinfo{author}{\bibfnamefont{D.}~\bibnamefont{Kleppner}}, \bibnamefont{and}
  \bibinfo{author}{\bibfnamefont{T.~J.} \bibnamefont{Greytak}},
  \bibinfo{journal}{Phys. Rev. A} \textbf{\bibinfo{volume}{81}},
  \bibinfo{pages}{062706} (\bibinfo{year}{2010}).

\bibitem[{\citenamefont{Hasted}(1972)}]{Hasted1972}
\bibinfo{author}{\bibfnamefont{J.}~\bibnamefont{Hasted}},
  \emph{\bibinfo{title}{{Physics of Atomic Collisions}}}
  (\bibinfo{publisher}{Butterworth \& Co Publishers Ltd},
  \bibinfo{address}{Washington, D.C.}, \bibinfo{year}{1972}).

\bibitem[{\citenamefont{Corliss and Bozman}(1962)}]{Corliss1962}
\bibinfo{author}{\bibfnamefont{C.}~\bibnamefont{Corliss}} \bibnamefont{and}
  \bibinfo{author}{\bibfnamefont{W.}~\bibnamefont{Bozman}},
  \emph{\bibinfo{title}{{NBS Monograph 53}}} (\bibinfo{address}{Washington,
  D.C.}, \bibinfo{year}{1962}).

\bibitem[{\citenamefont{Katz and Seaborg}(1957)}]{Katz1957}
\bibinfo{author}{\bibfnamefont{J.~J.} \bibnamefont{Katz}} \bibnamefont{and}
  \bibinfo{author}{\bibfnamefont{G.~T.} \bibnamefont{Seaborg}},
  \emph{\bibinfo{title}{{The Chemistry of the Actinide Elements}}}
  (\bibinfo{publisher}{John Wiley and Sons, Inc.}, \bibinfo{address}{New York},
  \bibinfo{year}{1957}).

\bibitem[{\citenamefont{Connolly}(2012)}]{Connolly2012}
\bibinfo{author}{\bibfnamefont{C.}~\bibnamefont{Connolly}},
  \bibinfo{type}{Ph.d. thesis}, \bibinfo{school}{Harvard University}
  (\bibinfo{year}{2012}).

\bibitem[{\citenamefont{Connolly
  et~al.}(2013{\natexlab{b}})\citenamefont{Connolly, Au, Chae, Tscherbul,
  Buchachenko, Ketterle, and Doyle}}]{Connolly2013b}
\bibinfo{author}{\bibfnamefont{C.~B.} \bibnamefont{Connolly}},
  \bibinfo{author}{\bibfnamefont{Y.~S.} \bibnamefont{Au}},
  \bibinfo{author}{\bibfnamefont{E.}~\bibnamefont{Chae}},
  \bibinfo{author}{\bibfnamefont{T.~V.} \bibnamefont{Tscherbul}},
  \bibinfo{author}{\bibfnamefont{A.~A.} \bibnamefont{Buchachenko}},
  \bibinfo{author}{\bibfnamefont{W.}~\bibnamefont{Ketterle}}, \bibnamefont{and}
  \bibinfo{author}{\bibfnamefont{J.~M.} \bibnamefont{Doyle}},
  \bibinfo{journal}{Phys. Rev. A} \textbf{\bibinfo{volume}{88}},
  \bibinfo{pages}{012707} (\bibinfo{year}{2013}{\natexlab{b}}).

\bibitem[{\citenamefont{DeCarvalho and Doyle}(2004)}]{Decarvalho2004}
\bibinfo{author}{\bibfnamefont{R.}~\bibnamefont{DeCarvalho}} \bibnamefont{and}
  \bibinfo{author}{\bibfnamefont{J.}~\bibnamefont{Doyle}},
  \bibinfo{journal}{Phys. Rev. A} \textbf{\bibinfo{volume}{70}},
  \bibinfo{pages}{053409} (\bibinfo{year}{2004}).

\bibitem[{\citenamefont{Lu et~al.}(2011)\citenamefont{Lu, Rasmussen, Wright,
  Patterson, and Doyle}}]{Lu2011}
\bibinfo{author}{\bibfnamefont{H.-I.} \bibnamefont{Lu}},
  \bibinfo{author}{\bibfnamefont{J.}~\bibnamefont{Rasmussen}},
  \bibinfo{author}{\bibfnamefont{M.~J.} \bibnamefont{Wright}},
  \bibinfo{author}{\bibfnamefont{D.}~\bibnamefont{Patterson}},
  \bibnamefont{and} \bibinfo{author}{\bibfnamefont{J.~M.} \bibnamefont{Doyle}},
  \bibinfo{journal}{Phys. Chem. Chem. Phys.} \textbf{\bibinfo{volume}{13}},
  \bibinfo{pages}{18986} (\bibinfo{year}{2011}).

\end{thebibliography}

\end{document}